\input  harvmac.tex
\overfullrule=0pt
\magnification 1200
\def\TeV{\rm TeV}
\def\gu{\gamma _{\mu}}
\def\GeV{\rm GeV}

\def\simge{\mathrel{%
   \rlap{\raise 0.511ex \hbox{$>$}}{\lower 0.511ex \hbox{$\sim$}}}}
\def\simle{\mathrel{
   \rlap{\raise 0.511ex \hbox{$<$}}{\lower 0.511ex \hbox{$\sim$}}}}
 
\def\slashchar#1{\setbox0=\hbox{$#1$}           
   \dimen0=\wd0                                 
   \setbox1=\hbox{/} \dimen1=\wd1               
   \ifdim\dimen0>\dimen1                        
      \rlap{\hbox to \dimen0{\hfil/\hfil}}      
      #1                                        
   \else                                        
      \rlap{\hbox to \dimen1{\hfil$#1$\hfil}}   
      /                                         
   \fi}                                         %

\def\CH{{\cal H}}

\def\ts{\thinspace}
\def\ra{\rightarrow}

\def\ra{\rightarrow}
\def\ol{\bar}

\def\GeV{{\rm GeV}}
\def\TeV{{\rm TeV}}

\def\condt{\langle \bar T T\rangle}

\def\condtij{\langle \bar T^i_L T^j_R\rangle}

\def\condtbt{\langle \bar t t\rangle}

\def\Dzero{{\rm D}\slashchar{\rm O}}

\Title{\vbox{\baselineskip12pt\hbox{BUHEP-96-29}}}
{\vbox{\centerline{ Top Decay in}
\vskip2pt
\centerline{Topcolor-Assisted Technicolor}}}

\centerline{Bhashyam Balaji}
\bigskip\centerline{Physics Department}
\centerline{Boston University}\centerline{Boston, MA 02215}
\vskip .3in   

\centerline{\bf Abstract}

Conventional technicolor models with light charged technipions 
($\pi ^{\pm} _T$)  lead
to  an unacceptably large contribution to $t\ra {\pi ^+_T} b$ decay rate.
Topcolor-assisted technicolor models also have additional PGBs called 
top-pions ($\pi ^{\pm}_t$) which  may 
contribute  to this decay. We study the potentially dangerous mixing of
charged top-pion and technipions in toy models of `natural' 
topcolor-assisted technicolor. 
We find that the $t\ra {\pi ^+_{t,T}} b$ decay rate in such models can
be within experimental limits 
due to a combination of heavy top-pion and  
small $\pi _t - \pi_T$ mixing.

\bigskip

\Date{10/22/96}

\vfil\eject

A natural, dynamical explanation for electroweak and flavor 
symmetry breaking is a
desirable alternative to the Higgs sector in the standard model of
electroweak interactions. In technicolor (TC) theories 
\ref\tcref{S.~Weinberg, Phys.~Rev.~{\bf D19}, 1277 (1979)\semi
L.~Susskind, Phys.~Rev.~{\bf D20}, 2619 (1979).},
electroweak symmetry breaking is accomplished by
the chiral symmetry breaking of technifermions which transform nontrivially 
under a new strong and unbroken gauge interaction
called technicolor. This 
yields the right masses of the weak gauge bosons when the characteristic
energy scale of technicolor interactions 
is about a TeV. In order to give masses to the fermions
without using fundamental scalars, one invokes an  
additional, spontaneously broken, gauge interaction  called extended
technicolor (ETC) 
\ref\sdlsetc{S.~Dimopoulos and L.~Susskind, Nucl.~Phys.~{\bf B155}, 237
(1979).},
\ref\eekletc{E.~Eichten and K.~Lane, Phys.~Lett.~{\bf 90B}, 125
(1980).}.

Experimental constraints from flavor changing neutral 
currents \eekletc \ and  the value of the S parameter
\ref\pestac{M.~Peskin and T.~Takeuchi, Phys.~Rev.~Lett.~{\bf 65}, 964
  (1990); Phys.~Rev.~{\bf D46}, 381 (1992); 
B.~Holdom and J.~Terning, Phys.~Lett.~{\bf B247}, 88 (1990)\semi
T.~Appelquist and J. Terning, Phys.~Lett.~{\bf B315} 139 (1993);
hep-ph/9305258.}
seem to suggest that technicolor is a walking
\ref\wtc{B.~Holdom, Phys.~Rev.~{\bf D24}, 1441 (1981);
Phys.~Lett.~{\bf 150B}, 301 (1985); 
T.~Appelquist, D.~Karabali and L.~C.~R. Wijewardhana,
Phys.~Rev.~Lett.~{\bf 57}, 957 (1986)\semi
T.~Appelquist and L.~C.~R.~Wijewardhana, Phys.~Rev.~{\bf D36}, 568
(1987)\semi
K.~Yamawaki, M.~Bando and K.~Matumoto, Phys.~Rev.~Lett.~{\bf 56}, 1335
(1986) \semi
T.~Akiba and T.~Yanagida, Phys.~Lett.~{\bf 169B}, 432 (1986).}
gauge theory.
In addition, ETC interactions seem to be inadequate
to account for the extremely large top quark mass
\ref\setc{T.~Appelquist, M.~B.~Einhorn, T.~Takeuchi and
L.~C.~R.~Wijewardhana, Phys.~Lett.~{\bf 220B}, 223 (1989);
T.~Takeuchi, Phys.~Rev.~{\bf D40}, 2697 (1989)\semi
V.A.~Miransky and K.~Yamawaki, Mod.~Phys.~Lett.~{\bf A4},129 (1989)\semi
K.~Matumoto, Prog.~Theor.~Phys.~{\bf 81}, 277 (1989) \semi
R.~S.~Chivukula, A.~G.~Cohen and K.~Lane, Nucl.~Phys.~{\bf B343}, 554
(1990).},
\ref\CDFD{F.~Abe, et al., The CDF Collaboration, Phys.~Rev.~Lett.~
    {\bf 73}, 225 (1994); Phys.~Rev.~{\bf D50}, 2966 (1994); 
    Phys.~Rev.~Lett.~{\bf 74}, 2626 (1995)\semi
    S.~Abachi, et al., The $\Dzero $ Collaboration, Phys.~Rev.~Lett.~{\bf
      74}, 2632 (1995).}.

Topcolor-assisted technicolor (TC2)
\ref\hill{C.T.~Hill, Phys.~Lett.~{\bf B345}, 483 (1995);hep-ph/9411426.}
is a recent attempt to 
address the unsatisfactory features of the technicolor scenario. 
The basic idea is that of generation sensitive gauge 
group replication. In the simplest version of TC2, the third generation is 
assumed to transform with the usual quantum numbers
under strong $SU(3)_1\times U(1)_1$ while the lighter generations transform
identically under a different (and weaker) group $SU(3)_2\times U(1)_2$.
At scales of about 1 TeV, $SU(3)_1\times SU(3)_2$ and $U(1)_1\times
U(1)_2$ are spontaneously broken to ordinary color (SU(3)) and weak 
hypercharge, respectively.
Electroweak symmetry breaking is still driven primarily by technicolor
interactions. In addition, the topcolor interactions (felt only by the
third generation quarks) with a scale near $1$ ${\rm TeV}$ generate 
$\condtbt$ and the very large top-quark mass. The ETC interactions 
are still required to  
generate the light fermion masses and a 
small but important 
contribution to the mass of the top quark ($m^{ETC}_t$). The 
reason for a nonzero $m^{ETC}_t$ is to give mass to the  
Goldstone bosons of t, b chiral symmetry breaking (top-pions).

As was pointed out by Chivukula, Dobrescu and Terning
\ref\cdt{R.~S.~Chivukula, B.~A.~Dobrescu and J.~Terning, 
Phys.~Lett.~{\bf B353}, 289 (1995);hep-ph/9503203.},
generic TC2 models suffer from a `$(\rho -1)$/naturalness' 
problem, where  $\rho = M^2_W/M^2_Z cos^2\theta _W$.
In other words, it may be 
unnatural to have $\rho - 1$ 
to be within experimental limits. 
This is because $U(1)_1$ (like topcolor $SU(3)_1$)
is expected to be 
strong  so that the top interactions are critical while the bottom
interactions are sub-critical. Then, the technifermion doublet 
responsible for the top and bottom  
ETC masses  has
custodial isospin violating  $U(1)_1$ 
couplings (and unacceptably large $\rho -1$)
even when the 
technifermions are degenerate.
A small $U(1)_1$ coupling 
requires the topcolor couplings to be unnaturally fine tuned 
for top (but not bottom) condensation.

Natural TC2 was introduced by Lane and Eichten
\ref\kleentc{K.~Lane and E.~Eichten, Phys.~Lett.~{\bf B352} 382 (1995);
 hep-ph/9503433.}
to address the $(\rho -1)$/
naturalness problem. They employ two 
different technidoublets 
for bottom and top ETC masses, an additional
doublet for the lighter generations, and  thereby 
transfer the isospin 
violating interactions to the weak $U(1)_2$.
The model has 
no gauge anomalies. The ETC gauge group is unspecified; instead 
nonrenormalizable operators allowed after imposing constraints (such as 
$U(1)_{1,2}$ symmetry, gauge anomaly 
cancellation and desired  intragenerational 
mixing pattern) are listed.
Subsequently, a more ambitious model (with colored technifermions)
was  developed by Lane 
\ref\klsyge{K.~Lane, BUHEP--96--2; hep-ph/9602221.} 
to explain topcolor breaking and the observed magnitude of 
generational mixing in TC2 models.

In this note, we shall discuss top decay in TC2 models. 
A light charged technipion in  conventional ETC 
models can be ruled out because of the large 
$t\rightarrow {\pi _T}b$ decay rate. This is due to the large coupling 
$m_t/F_{\pi _T}$, where $F_{\pi _T} = v/\sqrt{N_D}$, 
$v = 246$ GeV and $N_D$ is the number of techni-doublets. The decay 
rate $\Gamma (t\ra \pi ^+_T b)$ is then given by
\eqn\tepbde{\eqalign{
& \Gamma (t\ra \pi^+_T b) = {1\over {16\pi}} 
{\left(m^2_t-m^2_{\pi _T}\right)^2\over {F^2_{\pi _T} m_T}},\cr}}
where $m_{\pi _T}$ is the technipion mass.
The branching ratio of top to bottom quark and W is measured to be 
B($t\ra Wb) = 0.87^{+0.13}_{-0.30}(stat)^{+0.13}_{-0.11} (syst)$
\ref\inca{J.~Incandela, Proceedings of the 10th Topical Workshop on Proton--
Antiproton Collider Physics, Fermilab, edited R.~Raja and J.~Yoh, 
p.256(1995).}
. The standard model value for $\Gamma(t\ra W^+b)$ is 1.6 GeV with an
essentially $100\% $ branching ratio. The  bounds on $F_{\pi }$ and 
$m_{\pi}$ are plotted in Figure 1 for different values of $B(t\ra W^+ b)$.
We see that small values of $F_{\pi _T}$ and $m_{\pi _T}$ are excluded.

However, in TC2 models, the contribution of a light 
technipion  contribution to top decay is small 
since the $t-{\pi _T}-b$ coupling is only
$m^{ETC}_t/F_{\pi _T}$
and $m^{ETC}_t/m_t$ is $0.01-0.1$\hill .
The top-pion, on the other hand, couples with strength $m^{dyn}_t/F_t$, 
which is large since $m^{dyn}_t\approx m_t$ and $F_t = 70$ GeV 
is the top-pion decay constant.  
The ETC interactions responsible for ordinary top quark
mass  induce mixing between top-pions and technipions. The resulting 
pseudo-Goldstone bosons (PGBs)
 can lead to an unacceptably large $t\ra \pi b$ decay rate if 
allowed by phase space and if the top-pion component in the mixed 
PGB is large (see, e.g., Ref.
\ref\klsing{K.~Lane, Phys.~Lett.~{\bf 357B}, 624 (1995); hep-ph/9507289\semi 
E.~Eichten, K.~Lane, BUHEP--96--9/FERMILAB--PUB--96/075--7; hep-ph/9607213.}).
Hence, this note will discuss the effect on top quark decay 
of the mixing of  the charged PGBs in the top doublet sector 
(top-pions) and the techniflavor sector (technipions) in natural TC2
models. We carried out a detailed analysis in the 
toy models of \kleentc . The study of top decay in 
these can give us an idea about what can happen
in more general TC2 models.

In \kleentc , the gauge group is $G_{TC}\times SU(3)_1\times U(1)_1
\times SU(3)_2\times U(1)_2\times
SU(2)_W$. There are three  doublets 
of techniquarks : $T^l$, $T^t$ and 
$T^b$, where $T = (U, D)$. 
The three technidoublets are assumed to
transform under the same complex irreducible representation of the
technicolor gauge group, $G_{TC}$. They are $SU(3)_{1,2}$ singlets; for
details on hypercharge assignments see \kleentc . Hence, the 
flavor symmetry group  (ignoring for the moment broken $U(1)_1$ and 
ETC interactions) in the techniflavor sector is 
$SU(6)_L\times SU(6)_R$. When TC interactions become strong, this 
is spontaneously broken to an $SU(6)$ subgroup. 
The flavour symmetry in the top sector is 
$SU(2)_L\times U(1)$ which breaks spontaneously to $U(1)_V$.
It is the charged Goldstone bosons that are of relevance here.
They obtain mass from ETC and $U(1)_1$ interactions.

The ETC gauge group is unspecified. Instead, the model 
is assumed to have certain 
ETC--generated four-fermion operators consistent with all 
gauge symmetries. Firstly, there are the ETC--generated
two-technifermion (2T) interactions required for quark hard 
mass generation. These are of the following form:
\eqn\qTTq{\eqalign{
&\CH_{\ol u_i u_j} = {g^2_{ETC} \over {M^2_{ETC}}} \ts \ol q^l_{iL}
\gamma^\mu T^l_L \ts \ol U^l_R \gamma_\mu u_{jR}  \ts + \ts\ts {\rm
  h.c.} \cr
&\CH_{\ol d_i d_j} = {g^2_{ETC} \over {M^2_{ETC}}} \ts \ol q^l_{iL}
\gamma^\mu T^l_L \ts \ol D^l_R \gamma_\mu d_{jR}  \ts + \ts\ts {\rm
  h.c.} \cr
&\CH_{\ol t t} = {g^2_{ETC} \over {M^2_{ETC}}} \ts \ol q^h_L \gamma^\mu
T^t_L
\ts \ol U^t_R \gamma_\mu t_R   \ts + \ts\ts {\rm h.c.} \cr
&\CH_{\ol b b} = {g^2_{ETC} \over {M^2_{ETC}}} \ts \ol q^h_L \gamma^\mu
T^b_L
\ts \ol D^b_R \gamma_\mu b_R \ts + \ts\ts {\rm h.c.} \cr}}
Here, $q^l_{iL} = (u_i, d_i)_L$ for $i=1,2$ stands for the two light 
doublets and $q^h_L = (t,b)_L$.
Also, $g_{ETC}$ and $M_{ETC}$ are 
generic ETC couplings and gauge
boson masses.
The model also has two sets of ETC--generated four--technifermion (4T)
interactions corresponding to the two allowed choices of technifermion 
$U(1)_1$ charges  
called  cases A and B in \kleentc . The 4T interactions in set 
A are  $\CH_{\ol l t \ol t b}$, $\CH_{\ol t b \ol b l}$,
$\CH_{\ol b l \ol l t}$,  $\CH_{\ol t l \ol l b}$, $\CH_{\ol t b \ol t
  b}$
and $\CH_{\rm diag}$, where, for example,
\eqn\opA{\eqalign{
&\CH_{\ol l t \ol t b}= {g^2_{ETC} \over {M^2_{ETC}}} \ts \ol T^l_L
 \gamma^\mu T^t_L \ts (a_{_U} \ol U^t_R \gamma_\mu U^b_R  + a_{_D} \ol
 D^t_R \gamma_\mu D^b_R)  \ts + \ts {\rm h.c.} \cr
&\CH_{{\rm diag}}= {g^2_{ETC} \over {M^2_{ETC}}} \ts \ol T^i_L
 \gamma^\mu T^i_L \ts (b_{_U} \ol U^j_R \gamma_\mu U^j_R  + b_{_D} \ol
 D^j_R \gamma_\mu D^j_R)  \ts . \cr }}
The constants $a_{_{U,D}}$,... stand for unknown ETC--model--dependent
factors and, in the diagonal interaction, $i,j = l,t,b$.
The allowed operators in Case B are $\CH_{\ol l t \ol t b}$, 
$\CH_{\ol l b \ol b t}$ and $\CH_{\rm diag}$. 
The operators in Eqs. (1) and (2) are renormalized at scale $M_{ETC}$.
In addition, 
there are also 4T $U(1)_1$ operators generated by $Z^{\prime }$ exchange
(expected to be comparable to 4T ETC operators because the $U(1)_1$
coupling is large) 
which are determined by the $U(1)_1$ charges in the two sets (see \kleentc).

In the presence of the broken ETC and $U(1)_1$ interactions, $\CH ^\prime $, 
the spontaneously broken chiral symmetries are also explicitly 
broken and the Goldstone bosons (except those responsible for W and Z masses)
become massive. For weak perturbations, the masses can be estimated 
using chiral perturbation theory
\ref\vacalign{R.~Dashen, Phys.~Rev.~{\bf D3}, 1879 (1971);
S.Weinberg, Phys.~Rev.~{\bf D13}, 974 (1976)\semi
E.~Eichten, K.~Lane and J.~Preskill, Phys.~Rev.~Lett.~{\bf 45}, 225
(1980)\semi
M.~Peskin, Nucl~Phys.~{\bf B175}, 197 (1980); J. Preskill, 
Nucl.~Phys.~{\bf B177}, 21 (1981); 
K.~Lane, Phys. Scr. {\bf 23} 1005 (1981).}.

We now state the values of the parameters used in our analysis.
Since there are three techni-doublets in this model, 
$F_{\pi_T} = 246{\GeV}/\sqrt{3}$.
The value of $F_t$ follows from the Pagels-Stokar formula 
in the fermion loop approximation 
\ref\bhl{W.~A.~Bardeen, C.~T.~Hill and M.~Lindner, Phys.~Rev.~{\bf
    D41}, 1647 (1990).}:
\eqn\pagsto{\eqalign{
& {{F^2_t\over 2} = {N_c\over 16\pi ^2} m^2_{t, dyn} 
{\left( \ln {\Lambda ^2_t\over m^2_{t, dyn}} + k\right)}} \cr}}
where $N_c = 3$ is the number of colors and 
$\langle 0\mid \bar q^h_L\gu  \tau ^a q^h_L\mid \pi _t^b\rangle = i F_t
p_{\mu}\delta ^{ab}$.
For $m_{t, dyn} \approx 167$ GeV, $k \approx 1$ and $\Lambda _t\approx $ 
$1.07$ TeV, $F_t $ is about $70$ ${\rm GeV}$. We estimate $\condtbt$ 
using the NJL approximation
\eqn\ttnjl{\eqalign{
& \condtbt = {{4 N_c}\over {16\pi ^2}} m_t \Lambda ^2_t .\cr}}
Comparing with the naive dimensional analysis 
\ref\diml{A.~Manohar and H.~Georgi, Nucl.~Phys.~{\bf B234}, 189 (1984)\semi
H.~Georgi and L.~Randall, {\it ibid}, {\bf B276}, 241 (1986).}
estimate $\condtbt = 4\pi \kappa _t F^3_t$, we obtain $\kappa _t\approx 3.5$.
The technifermion condensate  
$\langle \bar U^i_L U^i_R\rangle _{\Lambda _T}$ (no sum over 
techniflavor index i) 
is similarly estimated to be $2\pi \kappa _T F^3_T$, where $\kappa _T$
is $O(1)$.
The ETC-generated top quark mass is given by
\eqn\metce{\eqalign{
& m^{ETC}_t (M_{ETC}) 
= {g^2_{ETC}\over M^2_{ETC}} \langle \bar U^t U^t\rangle _{M_{ETC}}
\approx {g^2_{ETC}\over M^2_{ETC}} {\left(M_{ETC}\over \Lambda
  _T\right)}^{\gamma _m} 
\langle \bar U^t U^t\rangle _{\Lambda _T} \cr}}
where $\gamma _m$ is 
the anomalous dimension of $\bar U^t U^t$ and expected to be close to
1 in a walking gauge theory\wtc .
The ETC-generated top mass  renormalised at scale $\Lambda _T$ is 
given by
\eqn\metct{\eqalign{
& m^{ETC}_t (M_{\Lambda _T}) \approx 
m^{ETC}_t (M_{ETC}) {\left({g^2_3 (\Lambda _T)}\over {g^2_3 (M_{ETC})}\right)
}^{12\over 29} 
{\left({g^2_1 (\Lambda _T)}\over {g^2_1 (M_{ETC})}\right)}^{-3\over 20} \cr}}
For $M_{ETC} = 30$ $\TeV$ (appropriate for the third generation)
, $g^2_{ETC} (M_{ETC}) = 4\pi $, 
\foot{ The ETC coupling is expected to be large in a walking gauge theory
\ref\klrm{ K.~Lane and E.~Eichten, Phys.~Lett.~{\bf B222} (1989)274\semi
K.Lane and M.~V.~Ramana, Phys.~Rev.~{\bf D44} (1991) 2678. }.}
$g_3^2 (\Lambda _T) = g_1^2 (\Lambda_T)\approx 10$, $\gamma _m = 0.8$, 
and $\Lambda _T = 
4\pi F_{\pi _T}\approx 1.8$  TeV,  one obtains $m^{ETC}_t (\Lambda _T) = 
4.9 \kappa _T$ GeV. To complete this sample analysis, 
$\kappa _T \approx 1.7$ yields $m_t = m^{dyn}_t+m^{ETC}_t = 175$ GeV \CDFD. 

There are several qualitative comments we should make about the 
vacuum alignment studied here.
First, the 4T interactions 
and  $\CH _{\bar t t}$  are the only operators of importance
when studying vacuum alignment. 
In the extreme walking limit ($\gamma _m = 1$), the vacuum energy
contribution of the 4T piece is independent of the $M_{ETC}$.
Hence, if the coefficients of the 2T and 4T interactions are 
of the same order, the 2T contribution to the vacuum energy
is smaller than the 4T piece by a factor of $\condtbt _{ETC}
/\condt _{ETC} \approx 
{\kappa _t F^3_t \Lambda _t}/
\kappa _T F_{\pi _T}^3 M_{ETC}\approx 0.01$.
The mixing between top-pion and technipion is given by this ratio and
so is small.
In contrast, in QCD-like technicolor, the two pieces  are comparable,
and consequently  there will be significant mixing.
A crude estimate of the top-pion mass is $M^2_{\pi _t} \approx 
m^{ETC}_t\langle \bar t t\rangle /F^2_t \approx \kappa _t\ts  
4\pi \ts m^{ETC}_t F_t$, i.e., $M_{\pi _t}\approx 150$ GeV. 
The mass of the technipions are typically 
$M^2_{\pi_T}\approx (4\pi \ts g_{_{ETC}} \ts \kappa _T \ts F_{\pi _T}^2/
\Lambda _{TC})^2 \approx 
(250\ts \kappa _T \ts g_{_{ETC}})^2$ ${\rm GeV}^2$.  In a model with more 
techifermion doublets (such as in \klsyge ), the technipions 
are lighter and there is likely to be 
more mixing between the top-pion and the 
technipions.

The $t - \pi ^+_t - b$ coupling is \hill
\eqn\tpib{\eqalign{
& {\epsilon 
m_t^{dyn}\over {\sqrt2 F_t}}\left[\bar t (1-{\gamma }^5) b\pi _t^+ + 
h.c.\right],\cr}}
where $m^{dyn}_t$ is the dynamical top quark mass (167 GeV here) and 
$\epsilon $ is the top-pion component in the normalized technipion 
mass-eigenstate ( for instance, $\epsilon = 1$  in the absence of 2T ETC
interactions).
This modifies the decay rate of $t\ra \pi ^+_t b$ to  
\eqn\tpbde{\eqalign{
& \Gamma (t\ra \pi^+_t b) = {\mid \epsilon \mid ^2
\over {16\pi}} {\left(m^{dyn}_t\over m_t\right)^2}
{\left(m^2_t-m^2_{\pi _t}\right)^2\over {F^2_t m_t}}\cr}}
which depends sensitively on $m_{\pi _t}$.

We now carry out a numerical analysis. The condensate matrix
$\condtij \propto W_{ij}$, where $W_{ij}$ is an $SU(6)$ matrix,  
depends on the choice of the interactions and is determined numerically. 
The PGB mass-squared matrix is thus determined (using Dashen's formula 
\vacalign \  ) 
and the masses and the mixings are then obtained on diagonalization. 
The conservation of electric charge implies that  
$W_{ij}$ is block-diagonal, i.e., 
$W = W_u\oplus W_d$, where $W_u$ and $W_d$ are 
$3\times3$ condensate matrices in the up and down sectors respectively.
In the isospin symmetric case, either $W_u = W_d$ or $W_u = W_d^*$ 
(see \vacalign ). The masses of the PGBs and their mixings 
are different for the two cases, but that does not affect 
our general conclusions. 

The allowed set of 4T ETC interactions for case A are
$\CH_{\ol l t \ol t b}$, $\CH_{\ol t b \ol b l}$,
$\CH_{\ol b l \ol l t}$,  $\CH_{\ol t l \ol l b}$, $\CH_{\ol t b \ol t
  b}$ and $\CH_{\rm diag}$. 
Isospin symmetry implies 
$a_U = a_D$, $b_U = b_D$ etc. We studied the patterns of vacuum
alignment in the isospin limit for various values of 
$g_{ETC}^2/{4\pi}$ (chosen to be between $0.4$ and 
$1.0$) for the 4T ETC  
operators in case A  and including $\CH_{\bar t t}$. 
The scale of the ETC interactions is
taken to be 30 TeV, which is appropriate for the third generation. 
The  coefficients were chosen so that the vacuum was 
aligned non-trivially and not close to a symmetry limit. Also, there 
are no massless Goldstone bosons, other than the ones corresponding 
to the longitudinal components of $W^\pm$ and $Z^0$. 
We find generically that 
there is only one  PGB with mass less than $m_t$ with  
a typical coupling to the top-pion  of about $(0.4)\ts m^{dyn}_t/F_t$. 
The branching ratio B$(t\ra Wb)$ is then found to be about 0.6 (or more), 
which is  consistent with current experimental results. 
Note that $\Gamma (t\rightarrow \pi^+_t b)$ is a sensitive function of 
$m_{\pi _t}$ and $\epsilon$. 
The top-pion is found to have a mass in excess of 200 GeV. Walking has
been assumed (approximated here by assuming a constant $\gamma _m = 0.8$), 
which raises the values of the PGB masses. 

The isospin symmetric interactions chosen for case B included 4T
interaction $\CH_{\ol tb \ol tb}$, suggested by $U(1)_1$ interactions 
(apart from the 4T ETC interactions $\CH_{\ol lt \ol tb}$, $\CH_{\ol lb
\ol bt}$ and $\CH_{diag}$ and $\CH_{\ol t t}$) so as  to make 
vacuum alignment non-trivial and break all chiral symmetries. 
In this case, we find that the generic situation is that 
there are two light PGBs (since some $g^2_{U_(1)}$ are negative)
with masses less than $m_t$ and 
with the magnitude of $\epsilon$ typically of $O(0.10)$. The top-pion is found 
to be heavier than the top quark. Here we find that B$(t\ra Wb)\ge 0.7$,
which is also
consistent with the experimental result stated above. 
The contribution to  $\Gamma (t\ra \pi^+_T b) $ due to the technipion 
component is small as $m^{ETC}_t/m^{dyn}_t \sim 0.05$ and can hence be ignored.

In conclusion, we find that the potential problems associated with top 
decay into
the  light PGBs in the toy models of natural TC2 studied here can be
resolved if technicolor is a walking gauge theory. This is because
the top-pion is then generally heavier than the top quark and the top-pion   
does not mix significantly enough with lighter technipion(s) to cause an 
unacceptably large top decay rate.
In a more elaborate model (such as in \klsyge ),
the PGBs are expected to be lighter than in the models studied
here and the technipions are expected to mix significantly with
the top-pion, because of the smaller value of $F_{\pi_T}$. A 
significantly better 
experimental determination of the Br$(t\ra Wb)$ would 
severely restrict the allowed parameter space in such models. 

\bigskip

\vfil\eject

{\bf Acknowledgements}

The author is grateful to E.~Eichten for providing a program to solve
the vacuum alignment program.
The author thanks K.~Lane for suggesting  and guiding 
the investigation based on the concern originally raised by
C.~T.~Hill. The author also thanks Sekhar Chivukula for comments on the
manuscript. This research was supported by the Department of Energy
under Grant No. DE-FG02-91ER40676. 

\bigskip
\bigskip
\bigskip

\centerline{\bf Figure Caption }

\bigskip

\item{[1]} The limits on the charged technipion mass as a function of
  $F_{\pi }$ from $B(t\ra W^+ b)$\ \inca.
  The curves (from left to right) correspond to $B(t\ra W^+ b)
  = 0.25$, 0.5 and 0.87 respectively. The excluded regions lie
  below the curves.

\listrefs

\vfil\eject

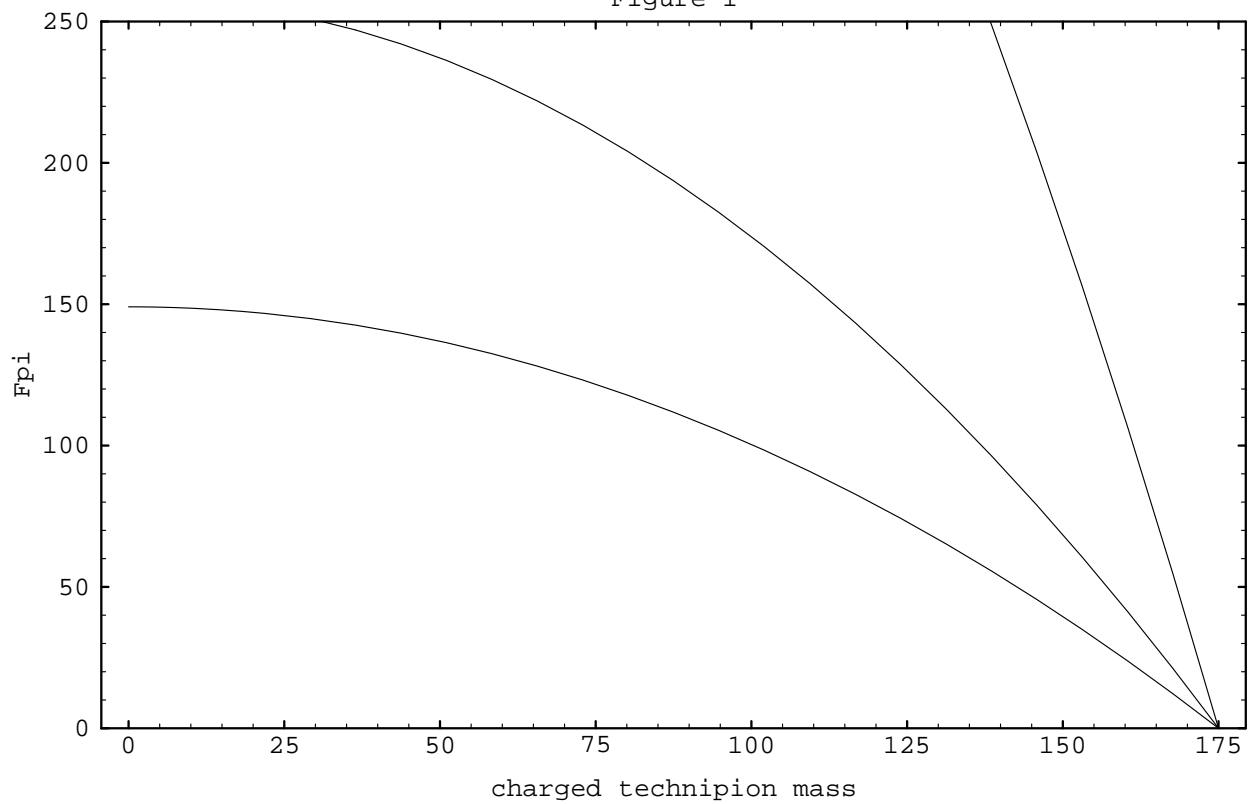
\bye